\documentstyle[preprint,prl,aps]{revtex}
\input{psfig}
\begin{document}
\tighten

\title{ Charm From Hyperons in the Future: Fermilab Experiment
781\thanks{Invited talk at Second International Workdshop on Heavy Quarks
Physics in Fixed Target,
University of Virginia, October 1994}
}
\author{
Michael Procario \\
{\em Carnegie Mellon University}
}
\preprint{\parbox[b]{1.5in}{CMU-HEP 95-01 \\ DOE-ER/40682-90} }
\maketitle
\baselineskip=14.5pt
\begin{abstract}
Recent results from CERN experiment WA89 have shown that charmed baryons and
particularly charmed-strange baryons have a significant cross section in
$\Sigma^-$ beams. Fermilab experiment 781, which is currently under
construction,  will utilize this fact to pursue a broad program of high
statistics studies of charmed baryons in the next Fermilab fixed target run.
\end{abstract}
\baselineskip=17pt
\section{Introduction}
\par\indent

As the previous speaker has clearly shown, charmed baryons and particularly
charmed-strange baryons are copiously produced in $\Sigma^- $ beams at $x_F >
0.2$. CERN experiment WA89 already has a sample of reconstructed
charmed-strange baryons that is is as good as any in the world, despite having
a relatively small sample of charmed mesons.  Fermilab experiment
781\cite{E781} will pursue this technique for studying charmed baryons with
higher beam fluxes, larger acceptance, and a better vertex detector.  This
will make E781 a second generation charmed baryon experiment, and allow the
systematic, in-depth study of charmed baryon production and decay physics.

The are a variety of physics goals in charmed baryon studies. The first is a
complete understanding of the weak decays. Charmed baryons like charmed mesons
have large QCD effects in their weak decays, and to understand these effects
will require a broad program of measurements. The first measurements should be
precision lifetimes of all weakly decaying charmed baryons. Today, only the
$\Lambda_c^+ $ is measured to better than 10\%. Bigi has called for all weakly
decaying charmed baryon lifetimes to be measured to better than
10\%.\cite{BIGI}

A very important complement to lifetimes are the semileptonic decays.
Currently, CLEO and ARGUS have evidence for $\Lambda_c^+ \rightarrow
X\Lambda\ell\nu $ and $\Xi^0_c \rightarrow X\Xi^-\ell\nu $ decays and CLEO
also has observed $\Xi^+_c \rightarrow X\Xi^0\ell\nu $ decays.\cite{semilep}
There is currently no evidence of a semileptonic decay where the final state
baryon is not a ground state hyperon.

Non-leptonic decays will be needed to sort out all of the nonfactorizing
effects that are expected in charmed baryons. Only in the case of the
$\Lambda_c^+ $ have significant non-leptonic decays been observed. Many of
these are multibody decays which may have resonance structure, but very little
is known about that now.

The second major area of physics that can be addressed with charmed baryons is
spectroscopy. Heavy Quark Effective Theory (HQET) has made a variety of
predictions about the spectrum of charmed mesons, and similar predictions about
baryons should be possible. There are extra degrees of freedom in the baryon
sector which can be used to more stringently test the potential models,
lattice calculations, and HQET predictions.  Experimentally, most of the
spectrum is unknown. Three of the four the ground state baryons are firmly
established, as are the isospin 1 triplet of $\Sigma_c $'s. New results on the
$\Omega_c $\cite{omegac}, excited $\Lambda_c$'s\cite{lamcstar}, and the
$\Xi_c^\prime $ are statistically
limited and will need to followed up. The rest of the spectrum is unknown.

The simple predictions of charm production from perturbative QCD do not agree
with the data, although these predictions have been successfully corrected by
modeling the non-perturbative effects.  Most of this work has been done for
mesons where the hadronization of the quark into the meson has been accounted
for with fragmentation models. Currently, the production of charmed baryons
seems to pose many more puzzles than have been seen in the mesons. NA32 has
reported observing equal rates of $\Lambda_c$ and $\bar\Lambda_c $ at large
$x_F $ from a $\pi^- $ beam, which is hard to understand in terms of
perturbative QCD. WA89 has seen strong leading particle effects in a $\Sigma^-
$ beam. They observe a much larger $\Sigma_c^0(cdd) $ signal than
$\Sigma^++_c(cuu) $.\cite{WA89} Both particles decay to $\Lambda_c^+ $ and a
charged $\pi$ so their acceptances are similar. One thing that is different is
that $\Sigma_c^0(cdd) $ has the same light diquark $(dd)$ as the $\Sigma^- $.

\section{The Detector}
\begin{figure}
\centerline{
\psfig{bbllx=30pt,bblly=570pt,bburx=580pt,bbury=770pt,file=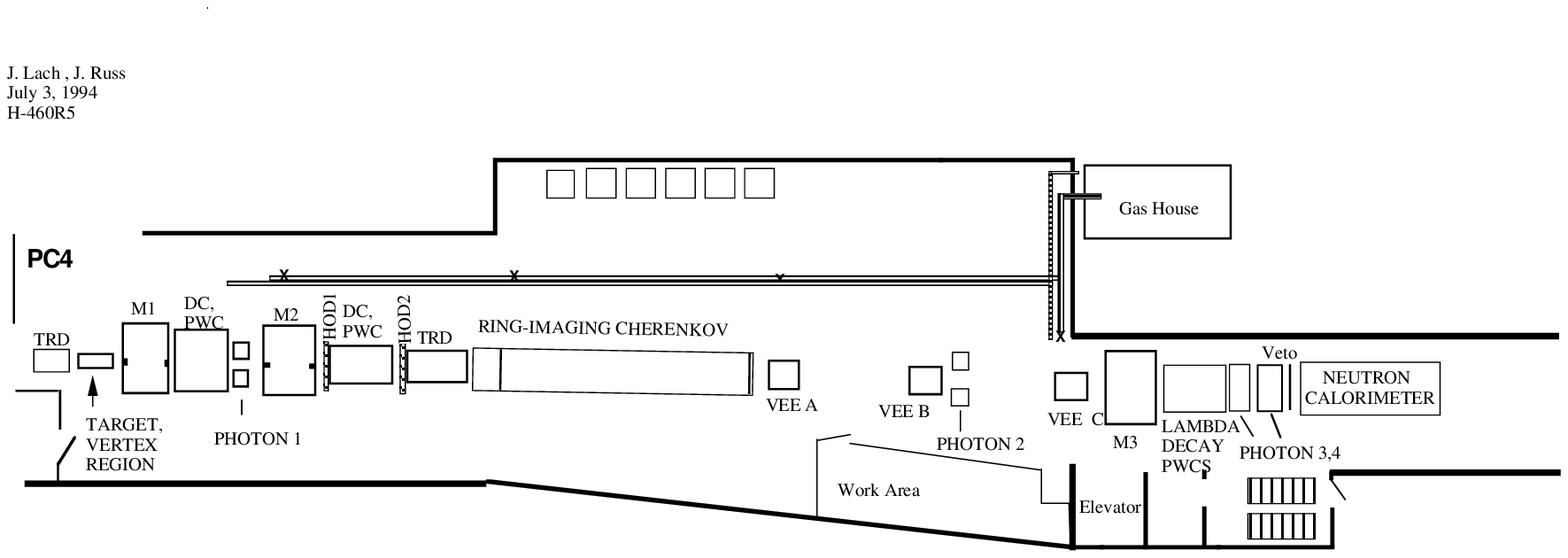,width=6.0in}
}
\caption{The layout of E781}
\label{layout}
\end{figure}

E781 is a three-stage forward charge-particle spectrometer with particle
identification and electromagnetic calorimetry.  The detector has acceptance
of $0.1 < x_F < 1.0 $. The overall layout of the detector is shown in figure
\ref{layout}. The are a variety of reasons for choosing this
geometry.

\begin{itemize}
\item At high $x_F $ the tracks have higher momentum and lower multiple
scattering. This improves the vertex resolution, and allows us to trigger on
large miss distance tracks.

\item For the high momentum tracks we have a small solid angle to cover. This
allows the use of a RICH with phototubes as the photon detector. Phototubes
are easier to use and build than to TAMI or CsI photocathodes.

\item It has been previously measured that the ratio of baryons to mesons
increases with x for strange particles, and the recent WA89 results have shown
the same effect for charmed particles.

\end{itemize}

The philosophy of the detector's design could be stated as: We know that
there is charm there and we should optimize on signal/background not signal.

\begin{figure}
\centerline{
\psfig{bbllx=90pt,bblly=110pt,bburx=690pt,bbury=480pt,file=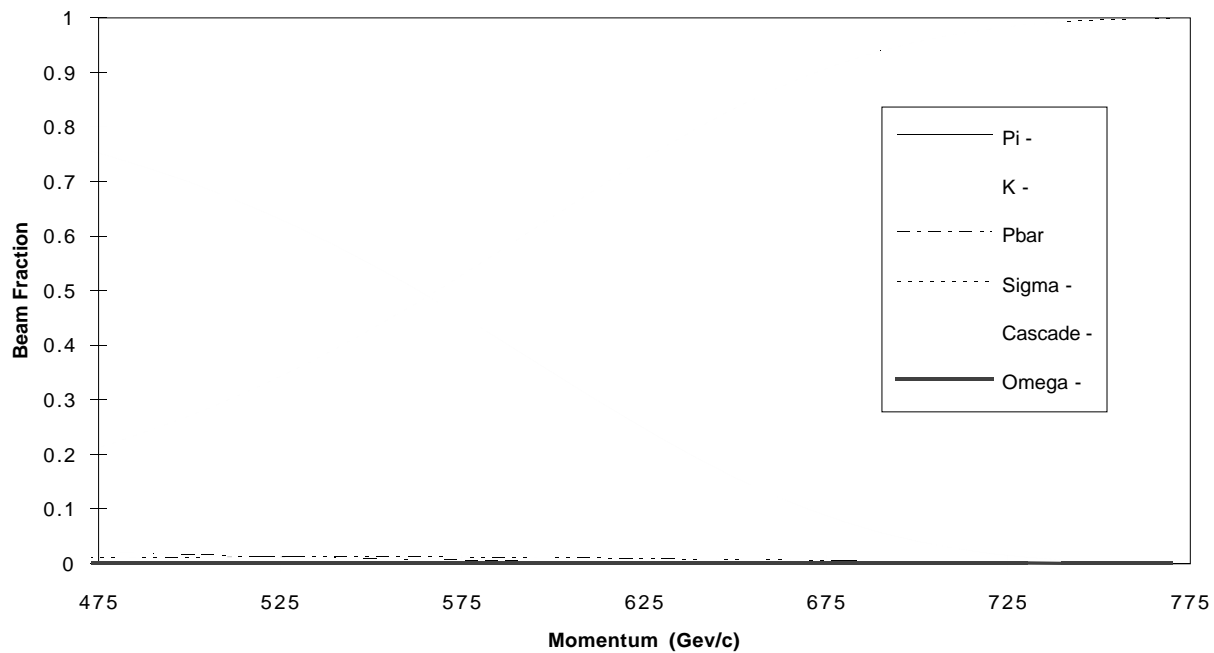,width=6.0in}
}
\caption{Particle fractions in Fermilab proton center hyperon beam}
\label{beam}
\end{figure}

The beam is predominantly mixture of $\pi^-$ and $\Sigma^-$ with a small
admixture of $\Xi^-$ and $\Omega^-$. The ratio, $n(\Sigma^-)/n(\pi^-)$ can be
adjusted by varying the momentum that is accepted as shown in figure
\ref{beam}. We plan to run with a ratio $n(\Sigma^-)/n(\pi^-) \ge 1$. The
$\Sigma $ flux will be $10^6$ MHz.

The first stage of the spectrometer has large acceptance with a 2.5 GeV/c
momentum cutoff. This stage measures soft pions from $D^*$'s, $\Sigma_c $'s
and other decays of excited charm states. It also can measure the tracks from
the other charm particle which is produced at lower $x_F$ than the trigger
charm particle, so that we can study charm pairs.

The second stage of the spectrometer has a 15 GeV/c momentum cutoff. This
stage is used for the trigger, which will be fully discussed in the next
section. There is a RICH detector with useful $p/K $ separation from 20 GeV/c
to 225 GeV/c and $K/p $ separation from 40 GeV/c to 480 GeV/c. There is a
transition radiation detector for electron identification.

The last stage measures the decay products of $\Lambda $'s that decay very far
downstream. Charmed strange baryon decays can decay $\Xi^- $ and $\Omega^- $
which produce $\Lambda$'s  very far downstream. This last stage is needed to
achieve high efficiency for these decays.

The beam is measured with a silicon strip system to provide high accuracy
predictions of $x-y$ position the primary vertex. The vertex region also has a
silicon strip detector. This detector has 20 planes in 4 views to provide
highly redundant tracking information to simplify track-finding and
track-matching both in the online software filter and offline reconstruction.
The performance of an eight plane system using an earlier generation of VLSI
readout was run in a test beam. It achieved excellent hit resolution of
$4\mu$m for planes with $20\mu$m pitch by interpolating the charge deposited
in adjacent strips.

There are three lead glass photon detectors for the reconstruction of $\pi^0$s
and photons. One array of lead glass is associated with each stage of the
spectrometer. The most downstream array will detect the radiative decay
photons like $\Sigma^0 \rightarrow \Lambda\gamma $ or the not yet confirmed
$\Xi_c^\prime \rightarrow \Xi_c\gamma $.

\section{Charm Trigger}

The heart of E781 is the hardware trigger and online software filter. Both of
these processes rely on the fact that the multiplicity is low in the second
stage spectrometer and that these tracks are high enough momentum that they
are well measured. The typical multiplicity of non-charm events is 15 at the
primary vertex but only 5 in our second stage spectrometer.

Two scintillator hodoscopes combined with matrix logic can count the number,
measure the charge, and roughly estimate the momentum of tracks in the second
spectrometer. By requiring 3 positive tracks in the second spectrometer, the
hardware trigger rejects non-charm by a factor of 8-10. Typically the charmed
baryon will contribute 2 of these positive tracks and the underlying event
will contribute the other. Events passing this trigger are fully read out and
passed on the online software filter.

The software filter runs in real time and only those event passing the filter
are written to tape.  This greatly
reduces the offline analysis load after the experiment finishes its run, but
it is critical that the quality of data is closely monitored to insure that
data is not lost.

The software filter is topological looking for evidence of a secondary vertex.
The filter searches for tracks after the second magnet. Those that are found
are projected back into the vertex detector. Since the multiplicity is low
after the second magnet the track finding is simplified there, and by looking
only along the projected tracks in the vertex detector the track finding is
also simplified in the vertex detector. These tracks are compared with the
intersection of the beam track and the target foils. If any of the tracks miss
this intersection by a significant amount then there is evidence of a
secondary vertex.

The angular acceptance of the second stage of the spectrometer is 30 mrad and
the targets are at most 1.5 mm thick, so the worst case geometric effect is
$22\mu$m. Multiple scattering errors are minimized by using only high momentum
tracks. Simulations studies have shown that a $30\mu$m cut on the miss
distance keeps the non-charm background trigger rate below 1\%, if there are
no tracking error. The fake trigger rate will be dominated by tracking errors
not measurement errors.

\begin{figure}
\centerline{
\psfig{bbllx=130pt,bblly=190pt,bburx=560pt,bbury=510pt,file=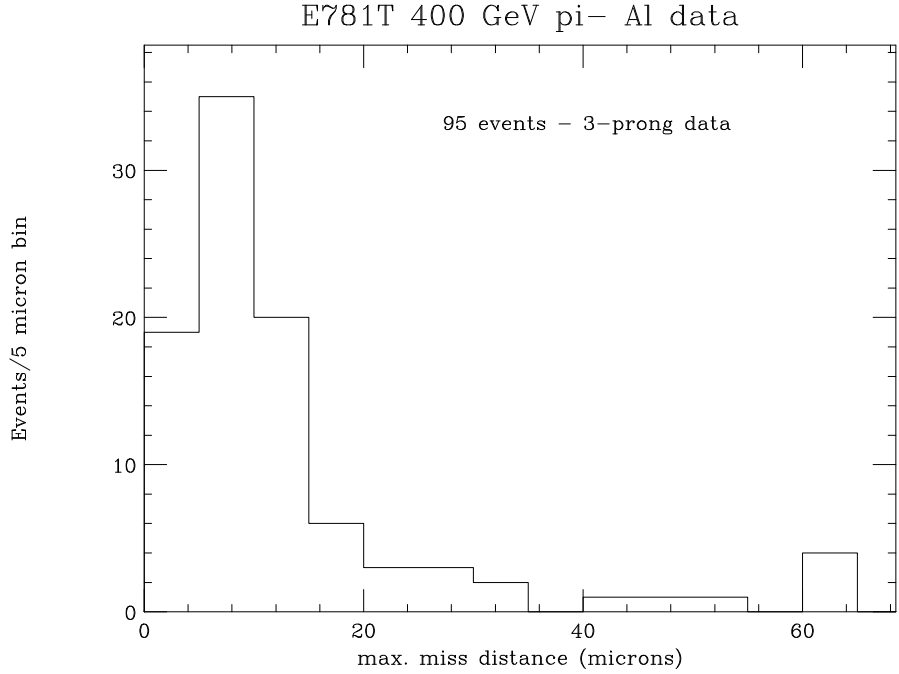,width=5.0in}
}
\caption{Test run results for maximum miss distance from primary of high
momentum tracks using an 8 plane silicon strip detector}
\label{testbeam}
\end{figure}

A test was performed with an eight plane silicon vertex detector and single
magnet spectrometer of similar angular acceptance as the full experiment. Data
was taken with a 400 GeV pion beam striking a thick target (6\% $\lambda_{int}
$ of Al).  Figure \ref{testbeam} shows the maximum miss distance per event
from the test run. The measured rejection was good, and the E781 trigger
should do better. A number of the events in the tail of the distribution had
hit confusion that will be helped by the stereo planes and the extra planes in
the full vertex detector.

The miss distance filter is fully efficient for charmed baryon decays with
lifetimes than 100 fs. It will also be very efficient for charmed meson decays,
since the filter requirement is just a secondary vertex.  The sample of charmed
mesons should be comparable to the sample of charmed baryons. It should be very
good for calibrating our detector, and we may be able to some small amount of
physics with it.

\begin{figure}
\centerline{
\psfig{bbllx=70pt,bblly=120pt,bburx=565pt,bbury=425pt,file=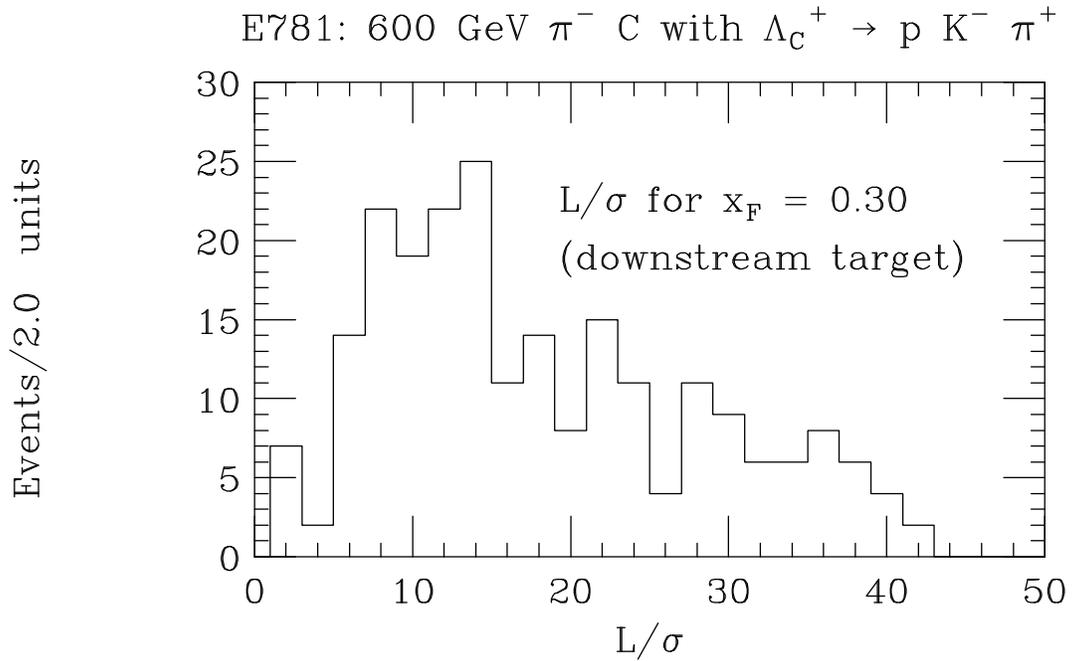,width=6.0in}
}
\caption{Monte Carlo results for the significance of $\Lambda_c^+ $ vertices
that pass the trigger}
\label{signif}
\end{figure}

Events which pass the miss distance filter are very useful for physics
analysis. The filter indirectly requires requires a secondary vertex. In most
analyses of charmed produced in fixed target experiments, the most powerful
rejection of background is achieved by requiring that the secondary vertex be
separated from primary vertex. This is usually expressed as the distance from
the primary vertex to the secondary divided by the resolution on the vertices
$(L/\sigma)$.  A simulation of $\Lambda_c^+ \rightarrow pK^-\pi^+ $ with
$x_F=0.3$ shows the effect of the miss distance trigger. Figure \ref{signif}
shows that events with low $L/\sigma $ have been removed, so they events that
pass our trigger are easier to analyze.

Some charmed baryons such as the $\Omega_c $ may have
shorter lifetimes.  We can also use other software filters designed around
different event characteristics. In $\Omega_c $ and $\Xi_c $ decays there is
multiple strangeness. Using the RICH to identify protons and kaons and select
events having both, and have sufficient rejection of background to not need
the miss distance requirement.

\section{Yields}

E781 plans to accumulate more than $10^6$ reconstructed charmed hadrons, with
over 100,000 in the large charmed baryon decay modes.  Using NA32 $\pi^- $
production cross sections\cite{NA32} we can predict what E781 can expect from
running with a $\pi^-$ beam. We scale up the cross section by 2 to account for
our higher energy beam. The $x_F $ distribution is different. The power is 4.2
instead of 3.5. We use the same $p_t $ spectrum.

The assumptions about the run are 1000 hours of data with 1000 seconds of
livetime per hour; 4\% interaction probability; and that charmed production
scales like $A^{1/3}$. The average $A$ of the E781 target is 32.8. The trigger
and reconstruction efficiencies have been calculated.  The trigger efficiency
weighted by the cross section
\[ \frac{d\sigma}{dx_F}=(1-x_F)^{4.2}.  \]
The reconstruction efficiencies were calculated using  all necessary effects,
such as detector resolution, multiple
Coulomb scattering,  primary and secondary vertex assignment, but not
pattern recognition mistakes. The results for $\pi^- $ data are shown in table
\ref{piminus-yield}.

The calculation of expected yields from the $\Sigma^- $ beam is more difficult
than for the $\pi^-$ beam because the WA89 cross section analysis is not
complete. We attempt to scale their yields by taking into account the relative
acceptances, rejection factors and number of triggered events. The assumptions
used are itemized below.
\begin{itemize}

\item The WA89 efficiency for $\Lambda_c \rightarrow pK^-\pi^+ $ is 1\% in
1991, 2.5\% in 1993 and 1994 for $x_F > 0.2 $.

\item WA89 trigger rejected inelastic events five times the rate it rejected
charm events.

\item WA89 uses a $L/\sigma $ cut of 5, which is similar to the E781 online
filter.

\item E781 will have 15 times more interactions.

\item E781 average efficiency per mode is 8\%.

\item The cross section at 600 GeV is 1.5 times greater than at 330 GeV.
\end{itemize}

Using these assumptions we arrive at the estimates in table \ref{sigma-yield}.
These estimates are only good to a factor of 2-3.

E781 will be able to take data with both beams simultaneously. This will
allow for systematic comparisons of the production from both beams.
Since the software filter gives us the ability to find charm while still
running, we will able to choose the beam that best optimizes our charmed baryon
yields.

\section{Conclusions}

Charmed baryon physics is maturing. Results are now coming in from a variety
of experiments on weak decays, spectroscopy and production mechanisms, which
is stimulating theoretical work in the is area. However, most of the current
results still have poor statistics compared to charmed mesons. The interesting
questions like the differences in charmed baryon lifetimes, the possible
leading particle effects, and many others will need higher statistics to be
answered.

An experiment optimized for the study of charmed baryons can significantly
improve this situation. E781 has set out to optimize the observation of
charmed baryons through the use of a $\Sigma^- $ beam, a very forward
geometery, excellent particle identification, and a topological trigger. The
yields expected in E781 will by on the order of 100,000 reconstructed charmed
hadrons in the large decay modes. This sample will have similar numbers of
charmed mesons and baryons and the charmed-strange baryons will be similar in
number as the charmed baryons.

\begin{table}
 \begin{center}
  \begin{tabular}{cccc} \hline
  Decay Mode & NA32 $\sigma\cdot B $ & E781 efficiency & Expected E781 yield \\
\hline
  $\Lambda_c^+ \rightarrow pK^-\pi^+ $ &
                $ 180 \pm 36 $      &     0.09        &    75,000         \\
  $\Xi_c^+ \rightarrow \Xi^-\pi^+\pi^+ $ &
                $ 130 \pm 95 $      &     0.06        &    40,000    \\
  $D^0 \rightarrow K^-\pi^+ $ &
                $ 230 \pm 40  $     &     0.08        &    86,000    \\ \hline
  \end{tabular}
 \end{center}
 \caption{ E781 anticipated charm yields from $\pi^- $ beam}
 \label{piminus-yield}
\end{table}

\begin{table}
 \begin{center}
 \begin{tabular}{cccc}\hline
 Decay Mode & WA89 (1991) & WA89 (1993) & Expected E781 yield  \\ \hline
 $\Lambda_c^+ \rightarrow pK^-\pi^+ $        &65 &$\sim 650$ & $\sim 50,000$ \\
 $\Xi_c^+ \rightarrow \Lambda K^-\pi^+\pi^+$ &42 &$\sim 400$ & $\sim 30,000$ \\
 $\Xi_c^0 \rightarrow \Lambda K^-\pi^+     $ &32 &$\sim 600$ & $\sim 50,000$ \\
 \hline
 \end{tabular}
 \end{center}
 \caption{ Estimates of expected charmed baryon yields from the $\Sigma^- $
beam in E781 scaled from WA89 yields}
 \label{sigma-yield}
\end{table}

\end{document}